\documentclass[%
pra,
 amsmath,amssymb,
 reprint,%
superscriptaddress
]{revtex4-1}

\usepackage{graphicx}
\usepackage{dcolumn}
\usepackage{bm}

\usepackage[utf8]{inputenc}
\usepackage[T1]{fontenc}
\usepackage{mathptmx}
\usepackage{etoolbox}
\usepackage{siunitx}
\usepackage{mathtools}

\makeatletter
\def\@email#1#2{%
 \endgroup
 \patchcmd{\titleblock@produce}
  {\frontmatter@RRAPformat}
  {\frontmatter@RRAPformat{\produce@RRAP{*#1\href{mailto:#2}{#2}}}\frontmatter@RRAPformat}
  {}{}
}%
\makeatother
\begin{document}

\preprint{AIP/123-QED}

\title[Title]{Measuring scattering distributions in scanning helium microscopy}

\author{C.J. Hatchwell}
\affiliation{Centre for Organic Electronics, University of Newcastle, Callaghan, NSW 2308, Australia.}
\author{M. Bergin}
\affiliation{Centre for Organic Electronics, University of Newcastle, Callaghan, NSW 2308, Australia.}
\email{matthew.bergin@newcastle.uon.edu}
 \author{B. Carr}
 \affiliation{Department of Physics, Cavendish Laboratory, JJ Thomson Avenue, University of Cambridge, Cambridge, CB3 0HE, UK}
\author{M.G. Barr}
\affiliation{Centre for Organic Electronics, University of Newcastle, Callaghan, NSW 2308, Australia.}
\author{A. Fahy}
\affiliation{Centre for Organic Electronics, University of Newcastle, Callaghan, NSW 2308, Australia.}
\author{P.C. Dastoor}
\affiliation{Centre for Organic Electronics, University of Newcastle, Callaghan, NSW 2308, Australia.}

\date{\today}

\begin{abstract}
A scanning helium microscope typically utilises a thermal energy helium atom beam, with an energy and wavelength (<100\,meV, $\sim$0.05\,nm) particularly sensitive to surface structure. An angular detector stage for a scanning helium microscope is presented that facilitates the in-situ measurement of scattering distributions from a sample. We begin by demonstrating typical elastic and inelastic scattering from ordered surfaces. We then go on to show the role of topography in diffuse scattering from disordered surfaces, observing deviations from simple cosine scattering. In total, these studies demonstrate the wealth of information that is encoded into the scattering distributions obtained with the technique.
\end{abstract}

\maketitle

%


\section{Introduction}

\label{sec:introduction}


A scanning helium microscope (SHeM) rasters a focused~\cite{koch_imaging_2008} or collimated beam~\cite{barr_design_2014,witham_simple_2011} of helium atoms across a sample, with measurements of the scattered flux being used to generate an image. 
The literature thus far has centred on exploring the macroscopic mechanisms underpinning available image contrast: occlusion (masking/shadowing) \cite{fahy_image_2018}, multiple scattering \cite{lambrick_ray_2018}, and diffuse illumination \cite{witham_exploring_2014}. Recent efforts have pushed beyond macroscopic contributions to contrast, demonstrating that the local order of the surface down to the angstrom scale directly affects the scattered intensity. Highly-ordered surfaces produce diffraction patterns~\cite{bergin_observation_2020,von_jeinsen_2d_nodate}, whilst the reflected intensity from disordered surfaces yields highly-specific topographic information (such as the growth modes of amorphous gold islands~\cite{eder_sub-resolution_2023}). 

The contrast observed in an image arises from the distribution of the atoms reflected from the sample~\cite{ginneken_diffuse_1998}. Therefore, furthering understanding of the mechanisms that underpin image contrast necessitates investigations of these scattering distributions. The scattering distribution arises from the averaging of individual scattering events within the illuminated area (spot size) on the sample surface. Feature sizes smaller than the lateral resolution of the instrument~\cite{bergin_standardizing_2022} - down to the level of surface defects and adsorbates - affect the resultant distribution. Typically, angularly resolved measurements of the scattering distribution have been achieved in helium atom scattering (HAS) by either rotating the sample relative to a stationary detector~\cite{head_specimen_1982}, or by moving the detector (either inside the vacuum chamber~\cite{dastoor_initial_1992}, or by moving the chamber itself~\cite{eder_focusing_2012}). While HAS systems offer excellent angular resolution ($\sim 0.1 ^\circ$), the spatial resolution is limited by the wide area of the sample illuminated by the helium beam ($\sim$\SI{1}{\milli\metre}$^2$). The spatial resolution limits the applicability of HAS to homogeneous surfaces possessing large-scale order~\cite{anemone_electronphonon_2019,anemone_electronphonon_2021,anemone_experimental_2020}, by necessity excluding anisotropic samples containing mixed domains of amorphous and crystalline structure. The pairing of neutral helium microscopy’s spatially-resolved capabilities with angularly-resolved measurements would offer a potent tool for surface physics investigations, to date no such instrument has been developed.

Advances in 3D printing have enabled the design of complex internal geometries to replicate the functionality of HAS within the confines of a SHeM~\cite{bergin_complex_2021}. Here, we present a design for an instrument that allows for the detection angle to be varied independently of the beam, while retaining the technique’s inherent spatially-resolved imaging capacity. The new apparatus is used to study four samples that illustrate a range of different mechanisms for forming contrast. We begin with investigating elastic diffraction from an ordered surface, then add complexity by looking at a sample that exhibits inelastic scattering and adsorbate coverage. We then study diffuse scattering from a known topography and finally present data on diffuse scattering from a technological surface.


\section{Instrumentation}
\label{sec:instrumentation}


\begin{figure}[t!]
\includegraphics[width=0.73\linewidth]{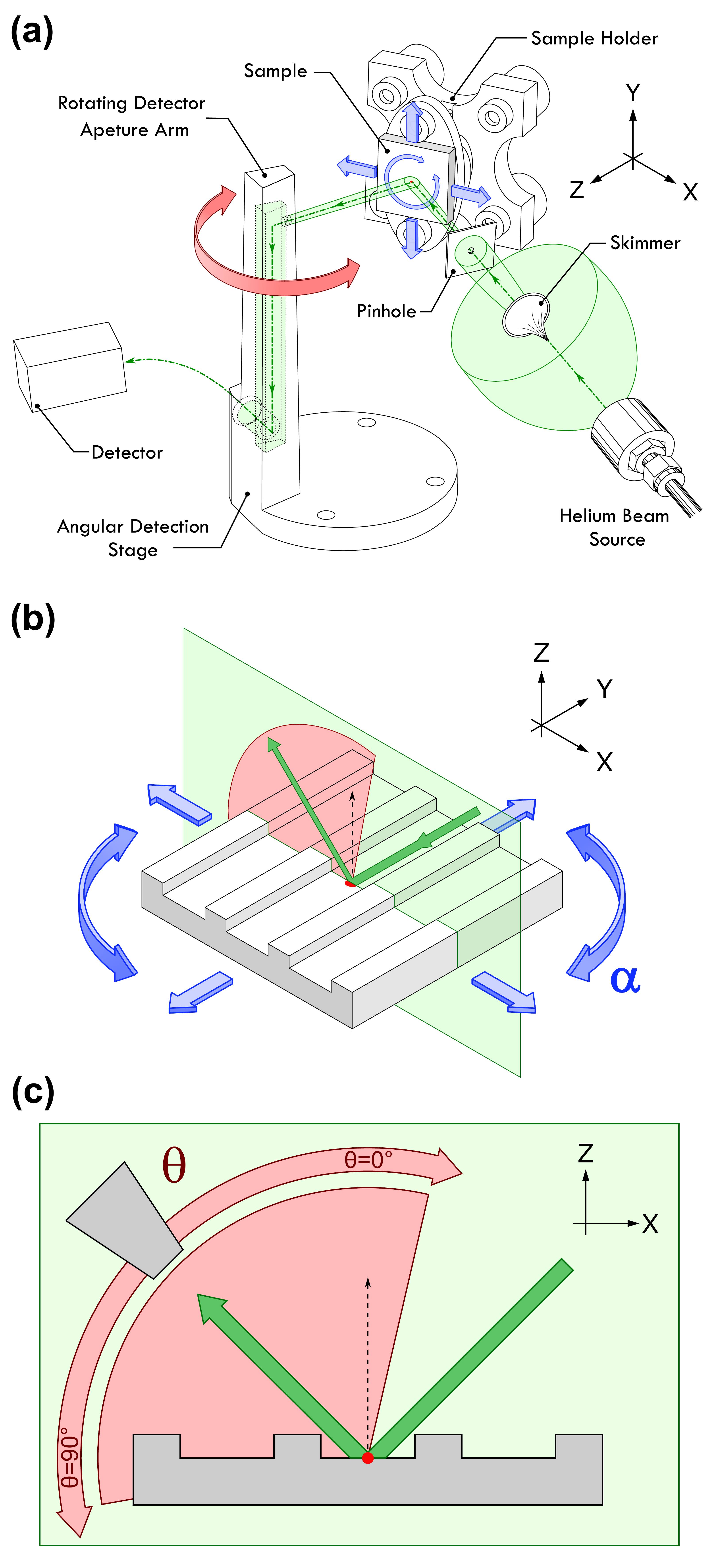}
\caption{(a) Schematic representation of the instrumentation changes, with the path of the helium beam through the optical elements shown in green. The new angular detection stage permits the rotation (represented by the red arrow) of the detector aperture attached to the end of a hollow arm about the coincidence point on the sample surface (red dot). Scattered helium from the sample is conducted through the arm and on to the primary stagnation volume of the detector via flexible tubing (not shown). The sample is mounted to a sample holder with temperature control and is capable of scanning in the XY-plane, with rotation about the Z axis (blue arrows). (b) Axonometric representation of the scattering of the incident neutral helium beam (green arrow) from the sample surface. Blue arrows show the possible translation of the sample in the XY-plane and rotation about the Z-axis ($\alpha$). The green plane (XZ) represents both the direction along which the beam is incident on the sample as well as the plane through which measurement takes place, with the detector aperture able to sweep through the light red arc shown centered on the coincidence point (red). (c) Side view of the scattering plane. The detector aperture (mounted to the rotating arm) is able to sample the scattering distribution across the red arc. Position of the aperture is given by $\theta$, with zero referenced to the sample normal (dotted arrow). The angular detection stage is typically capable of accessing $-13^\circ < \theta < 90^\circ$, with the latter limit being sample dependant.}
\label{fig:overview_im}
\end{figure}

Within the field of neutral atom microscopy, a scanning helium microscope (SHeM) refers to a specific type of instrument geometry comprising a supersonic free-jet expansion of helium, an optical element, a moving sample stage, and a neutral species detector. The optical element is usually housed within a modular pinhole plate, incorporating both a pinhole for beam collimation and a detector aperture for collection of the scattered flux~\cite{barr_design_2014,fahy_highly_2015}. The nature of the pinhole plate locks in the measurement geometry; typical designs have employed a 90-degree scattering angle with a 45-degree angle of incidence. Given that the detector aperture is fixed in place, such geometries preclude direct measurement of the entire scattering distribution, limiting all experiments to a restricted range of in-plane scattering angles. Initial attempts to subvert this limitation used a technique known as a ‘z-scan’~\cite{bergin_observation_2020}: the sample position is moved along the beam axis (adjusting the working distance), allowing the outgoing detection angle to vary. However, a z-scan also results in a changing detector solid angle as a function of working distance, the effect of which dominates the measured change in scattered intensity, complicating analysis. Other attempts to examine the scattering distribution in SHeM have involved probing scattered flux across an artificially-curved surface, which varies both the local incidence and detection angles simultaneously~\cite{lambrick_observation_2022}. This approach reveals fundamental information about the contrast mechanisms available to the technique, but is not applicable to general microscopy and surface analysis. 

Prior pinhole plate designs maintain a fixed geometry between the incident beam pinhole and the detection aperture, with the latter coupled to an ionisation detector operating in stagnation mode~\cite{barr_design_2014,fahy_highly_2015} (where the helium has scattered from the chamber walls to form a stagnant gas pressure). Stagnation mode removes the requirements for a direct line of sight between the sample and detector, allowing for the separation of the aperture and the primary stagnation volume. Provided that the conductance between the aperture and the stagnation volume is kept approximately constant, a short flexible hose may be used to connect the two. The flexibility of the connection means that the detector aperture can be rotated in a circular arc around the sample of interest, facilitating direct measurements of the scattering distribution. Taking advantage of this principle, a new `angular detection stage' was developed for a SHeM~\cite{fahy_highly_2015}. In addition, the existing sample holder (allowing for scanning underneath the beam and control of the working distance) was modified to incorporate temperature control and sample rotation. A schematic representation of the instrumentation changes to the SHeM are shown in Figure~\ref{fig:overview_im} (a).
Standard operating conditions (beam pressures, path lengths, dwell times, etc) have been described previously in the literature~\cite{fahy_highly_2015, eder_sub-resolution_2023}.

Ultra high vacuum (UHV) compatible components with complex geometries can now be produced at low cost using stereolithographic (SLA) 3D printing~\cite{bergin_complex_2021}. The components of the angular detection stage were fabricated via a Formlabs Form3B 3D printer from clear V4 resin, using the standard printing, cleaning and curing process~\cite{formlabs_guide_2022}. After curing, the components were baked at $110 ^\circ$C for 12 hours in a vacuum oven to remove excess water from the resin and allow compatibility with a UHV environment \cite{bergin_complex_2021}. Dymax 3220-GEL-SC UV-cure epoxy was used to attach the flexible silicone tubing ($\varnothing$ \SI{2}{\milli\metre} internal diameter, \SI{0.4}{\milli\metre} wall thickness) between the rotatable detector aperture and the primary stagnation volume. 

Rotation of the detector aperture was achieved though mounting to an Attocube ECR3030 (cat. no. 1006211) nanopositioning stage. An additional linear Attocube ECS3030 (cat. no. 1011434) nanopositioner was also installed to allow movement of the detector arm  in the X-axis (refer to Figure~\ref{fig:overview_im}) and correct any misalignment between the arm’s centre of rotation and the point where the helium beam scatters off the sample (the coincidence point). The existing sample holder for the SHeM utilised three Attocube ECS3030 nanopositioners in an orthogonal stack~\cite{myles_taxonomy_2019}. The holder was updated to facilitate rotation of the sample about the Z-axis, henceforth referred to as $\alpha$ (see Figure~\ref{fig:overview_im}), via a second Attocube ECR3030 unit mounted on the Y-axis of the orthogonal stack. Sample temperature control was realised via a UHV button heater (HeatWave Labs part no. 101136) and K-type thermocouple with PID control. A MACOR® insulator was incorporated to thermally-isolate the nanopositioners. To support the added mass, the Y-axis ECS3030 was provided with an Attocube weight compensator (ECS Lift 3030 Art. no 1011297). 

Combining the functionality of HAS within the physical geometry of a SHeM requires careful attention to the competing demands of spatial resolution, angular resolution, and imaging time \cite{myles_fast_2020,bergin_standardizing_2022}. With these factors in mind, a balance must be struck between the diameter of the pinhole and the detector apertures, as well as the arc subtended by the detector arm. Through testing multiple iterations of the stage – facilitated by 3D printing – a \SI{25}{\micro\metre} diameter pinhole, a \SI{0.5}{\milli\metre} diameter detector aperture, and a \SI{7}{\milli\metre} detector arm arc radius were chosen. This configuration yields an angular resolution of $4.1 ^\circ$, whilst maintaining imaging times comparable to the prior instrument setups~\cite{fahy_highly_2015, eder_sub-resolution_2023}. In practise, the additional path length and volume contributed by the flexible tubing was not found to present an appreciable restriction in overall imaging time. We define $\theta$ as the angle that the detector arm describes about the coincidence point, referenced to the sample normal ($\theta = 0$). The included angle is restricted to a maximum of $-13 ^\circ$ in the backscatter direction (toward the incident beam) by the physical proximity to the pinhole plate. In the forward scattering direction, $\theta$ is typically limited up to $90 ^\circ$, however this restriction is highly dependent on the size and shape of the sample under investigation. The described instrument configuration was employed for all experimental work appearing in this manuscript.

Accurate analysis of any collected data is reliant on the precise alignment of the angular detection stage. The centre of rotation of the detector arm must match the coincidence point to ensure that no change of the solid angle of the detector aperture occurs during measurement. Misalignment of these points results in changes in intensity not simply due to the scattering distribution as desired, but rather a convolution of the distribution and the instrument geometry. To achieve precise alignment, the length of a mask produced by a step feature of known height was measured; a full description of the procedure appears in appendix~\ref{sec:alignment}. 

The described angular detection stage retains the original imaging capabilities of the instrument~\cite{fahy_highly_2015, eder_sub-resolution_2023} while facilitating additional scan modes. Previous configurations were restricted to monitoring a single outgoing direction, but the rotating arm allows imaging with our choice of detector aperture position. Beyond imaging, by fixing the sample position and sweeping the detector aperture through the available values of $\theta$ we may investigate the scattering distribution for that point. Finally, the sample may also be rotated in $\alpha$ to alter the incoming azimuthal angle of the incident beam with respect to the surface under investigation. The remainder of this paper demonstrates the strength of these capabilities for either imaging or spectroscopic measurements.


\section{Application}

\subsection{Elastic scattering from an ordered surface}

\begin{figure}[t]
\includegraphics[width=0.95\linewidth]{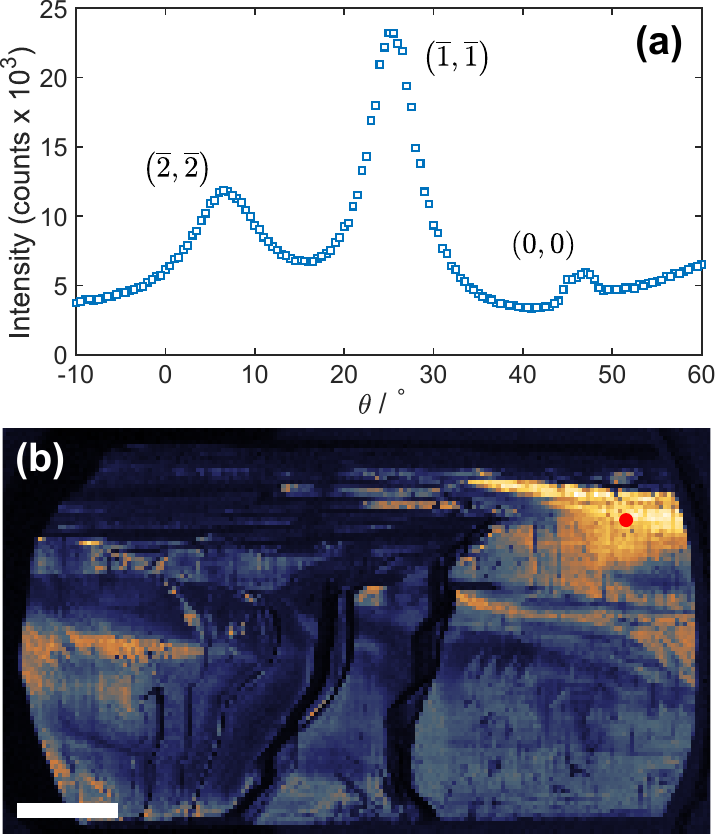}
\caption{(a) Angular scan taken from the LiF sample aligned on the <100> direction. The peaks demonstrate diffraction from the surface and are labelled $(hk)$ with respect to the cubic lattice of the fluoride ions. (b) Image of the LiF sample when positioned on the 1st order diffraction peak, the scalebar is 1\,mm. Variations in the surface topograhy are seen as the diffraction peak moves into and out of the detector or where no diffraction occurs.}
\label{fig:LiF}
\end{figure}

It is well known from HAS that ordered surfaces are capable of diffracting helium~\cite{farias_atomic_1998}, leading to structure in the scattering distribution. These samples must be homogeneous over $\sim$ 1\,mm$^2$ to be studied by the broad illumination typical of HAS. Lithium fluoride (LiF) has been studied extensively by atom scattering due to the unique properties of the crystal~\cite{estermann_beugung_1930, boato_study_1976, celli_pairwise_1985, riley_refined_2007,von_jeinsen_2d_nodate}. In the bulk, LiF has an f.c.c. crystal lattice, however the electron density above the surface is determined by the positions of the fluoride ions causing the (100) plane to possess a simple cubic structure. The electron density above the surface is strongly modulated across the atoms, leading to large diffraction peaks. The strongly bound atoms also lead to minimal inelastic scattering occurring and an inert surface that allows surface preparation by cleaving out of the vacuum.

For elastic scattering, the diffraction peaks will appear at
\begin{equation}
\mathbf{K_f}=\mathbf{K_i}+\mathbf{G},
\end{equation}
where $\mathbf{K_i}$ and $\mathbf{K_f}$ are the two dimensional wavevectors of the incoming and scattered helium (parallel to the surface) and $\mathbf{G}$ is a two dimensional surface reciprocal lattice vector.

\begin{figure*}[t]
\includegraphics[width=0.95\linewidth]{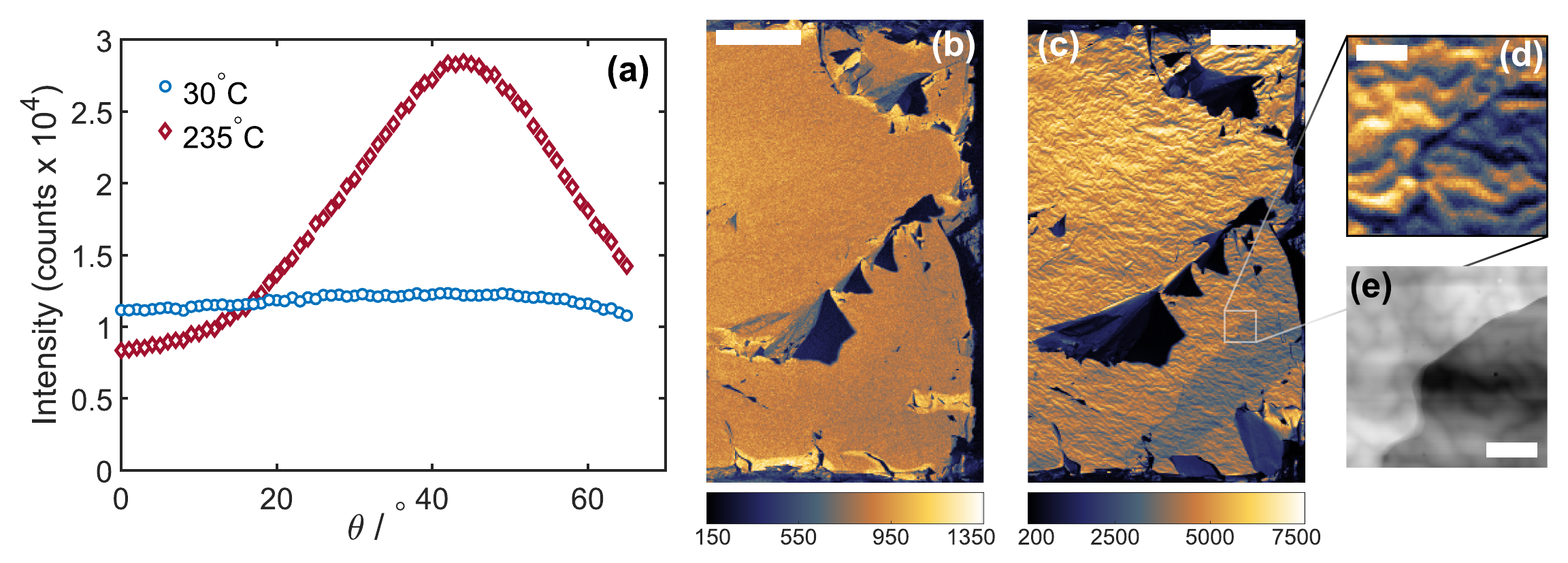}
\caption{(a) Comparison of angular scans taken from the same point on the HOPG sample at temperatures of 30°C (blue points) and 235°C (red points). After heating, a contamination layer of adsorbates is removed, revealing the scattering distribution of the bare HOPG. Micrographs of HOPG taken at sample temperatures of (b) 30°C and (c) 210°C; both scale bars  2mm. The additional contrast revealed at high temperature is due to the peaked nature of the scattering distribution. (d) High resolution helium micrograph collected at temperature from the indicated region in (c), with (e) a surface map of the same area acquired via a stylus profilometer. Both scale bars are \SI{100}{\micro\metre}. The topography observed in the profilometer map matches the contrast observed in the helium image, corroborating the  sensitivity of the probe particle to overlying adsorbate layers.}
\label{fig:HOPG}
\end{figure*}

The collimated beam in combination with the new angular detector stage allows the SHeM to collect diffraction patterns from any point across a heterogenous surface. Figure~\ref{fig:LiF}(a) shows the angular distribution obtained from a LiF crystal along the <100> azimuth as collected by the new detector stage. The specular peak appears as a top hat function near $\theta = 45^\circ$ due to the finite size of the detector aperture convoluting the sharp elastic peak. The additional peaks observed at $\theta < 45^\circ$ are the diffraction peaks from the surface lattice. These diffraction peaks appear broadened due to the velocity distribution of the incoming helium atoms.

Figure~\ref{fig:LiF}(b) is an image of the surface of the LiF crystal obtained when the detector is positioned to collect the first order diffraction peak. Note that figure~\ref{fig:LiF}(a) was collected at the point marked with a red circle. Any small changes in the topography of the surface will cause shifts in the local surface normal, therefore causing the diffraction peak to move slightly. The image shown in figure~\ref{fig:LiF}(b) maps both these small topographic changes and regions of the crystal where diffraction is not occurring. Therefore, the angular detector stage allows the scattering distribution to be mapped out at any point in the image due to the spatially resolved beam.

\subsection{Inelastic scattering from an ordered surface}

In contrast to LiF, not all surfaces can be represented with a perfectly periodic lattice without any adsorbates, nor will the scattering be purely elastic. These disturbances from the ideal scenario will cause the scattering distribution to deviate from matching the reciprocal lattice of the surface. ZYB grade highly oriented pyrolitic graphite (HOPG)~\footnote{From Ted Pella Product No. 626-10} demonstrates these additional complications and can be prepared out of the vacuum via the tape exfoliation process, then heated in-situ to remove adsorbates. Experiments at elevated temperatures typically involved first flashing the sample to 290°C for 5-10 minutes. However, given the sample chamber pressure is $\sim$\SI{5e-9}{\milli\bar}, the surface would fully recontaminate within 12 hours at room temperature. 

Figure~\ref{fig:HOPG}(a) compares scattering distributions from HOPG at ambient temperature and 235°C. We attribute the distinct change in the scattering distribution post-heating to the surface desorbing a contamination layer and revealing the ordered surface structure. The observed peaked scattering distribution when the surface is clean is consistent with prior measurements of thermal energy helium scattering from room temperature HOPG~\cite{oh_elastic_2009,shichibe_probing_2015}. Unlike with LiF, the properties of the top layer of HOPG (such as the low mass of the surface atoms) leads to the inelastic scattering channel dominating over the elastic channels~\cite{oh_elastic_2009}. We do not observe a sharp elastic specular peak as seen in the angular scan of LiF, due to either a small misalignment of the sample causing the specular peak appearing slightly out of plane~\cite{von_jeinsen_2d_nodate}, or the Debye-Waller attenuation~\cite{oh_elastic_2009,shichibe_probing_2015} of the peak meaning it cannot be observed above the inelastic feature.

Given the large helium scattering cross-section from defects and adsorbates~\cite{farias_atomic_1998}, it might be expected that the angular distribution measured from the HOPG surface at ambient temperatures in Figure~\ref{fig:HOPG}(a) would completely randomise the outgoing trajectories of the helium atoms and produce a cosine distribution. However, we note that while the distribution for the ambient temperature surface in Figure~\ref{fig:HOPG} (a) is less peaked than for the hot (clean) surface, the distribution still has a maximum near the specular direction, rather than at the surface normal at $\theta=0^\circ$. We therefore conclude that when the HOPG surface is covered in adsorbates, it is still not sufficiently rough to completely randomise the helium scattering.

Figure~\ref{fig:HOPG}(b) is a micrograph of the HOPG taken at an ambient 30°C and Figure~\ref{fig:HOPG}(c) has been heated to 210°C, where both images were collected at $\theta = 45^\circ$. The shift from a diffuse distribution to one with a strong angular dependence -- and a peak centered around the specular -- accounts for the vivid surface structure seen in Figure~\ref{fig:HOPG}(c). The increased angular dependence of the scattering distribution results in small changes in the local normal having a large effect on the measured signal, revealing the detailed surface topography. To investigate the origin of the image contrast, Figure~\ref{fig:HOPG}(d) shows a small area SHeM micrograph, while Figure~\ref{fig:HOPG}(e) displays an surface map produced using a stylus profilometer (Bruker Dektak XT) at the same area. The same structures are seen in both images, confirming that the structure originates from small scale variations in the surface topography.

\subsection{Diffuse scattering from a known topography}
\label{sec:characterisation}

\begin{figure}[t!]
\includegraphics[width=\columnwidth]{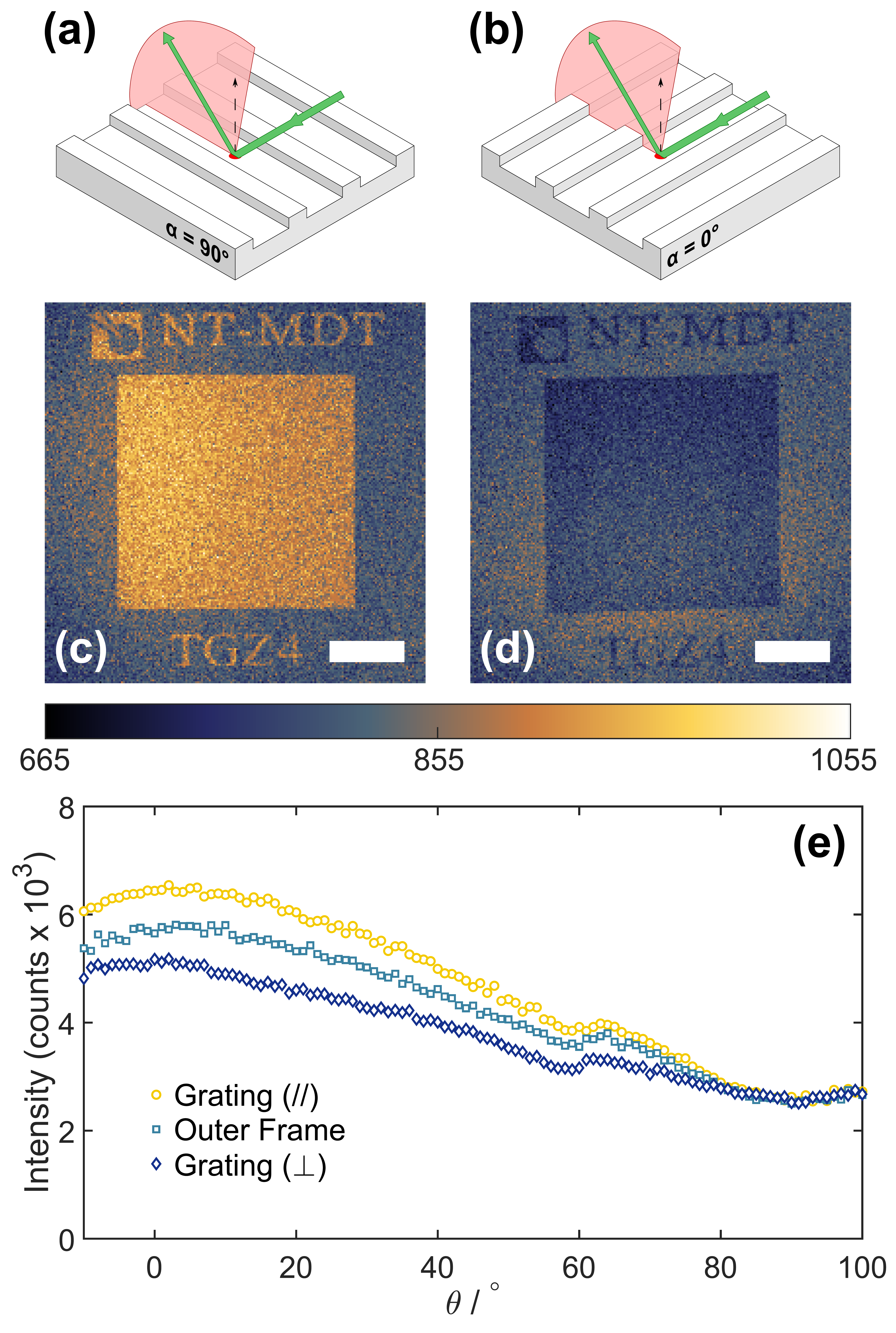} 
\caption{(a, b) Schematic representations of the helium beam interaction with an AFM grating at $\alpha$ = $\pm 90^\circ$ (parallel, $\parallel$, to the incidence angle of the beam) and $\alpha$ = $0^\circ$ (perpendicular, $\perp$, to the beam), respectively. (c, d) Neutral helium micrographs of the Tipsnano `TGZ4' AFM gratings (obtained at $\theta$ = $0^\circ$) with the grating parallel to and perpendicular (respectively) from the beam as represented in (a) and (b). While the scattered intensity from the (relatively) featureless silicon frame does not change between orientations, there is a contrast inversion on the grating. Scale bars for both micrographs \SI{1000}{\micro\metre}. (e) Plot of the scattering distribution as collected from the outer frame of the sample, as well as the grating in the parallel and perpendicular orientations. In the parallel orientation an enhanced intensity is observed relative to the outer frame, while in the perpendicular orientation the scattering distribution appears to have shifted towards the back scattering direction.
}
\label{fig:gratings_fig}
\end{figure}

\begin{figure}[t!]
\includegraphics[width=\columnwidth]{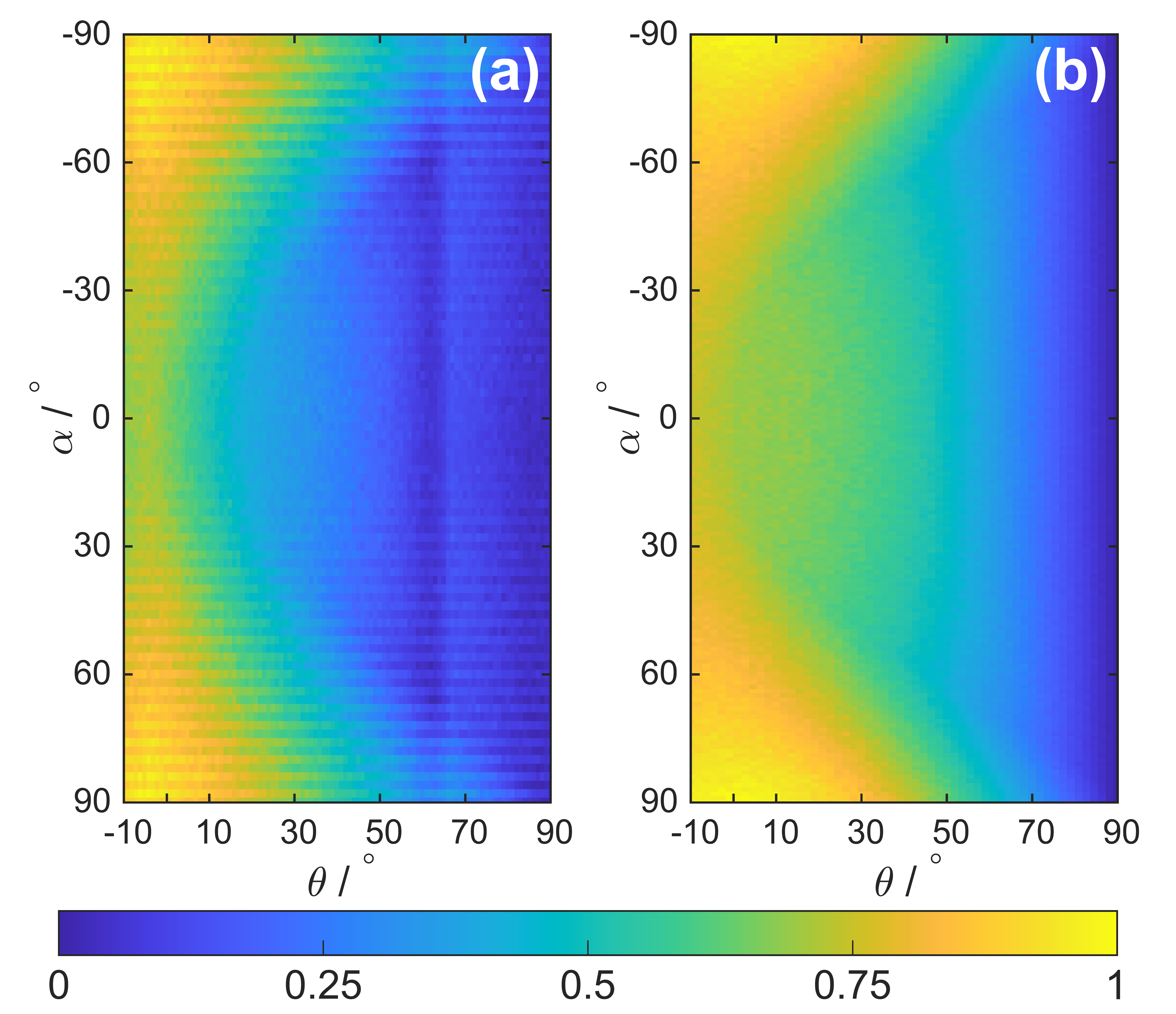} 
\caption{Comparison of the results of a ray-tracing simulation with scattering distributions as collected by the angular detection stage. (a) Plot of the normalised helium intensity as collected from the TGZ4 grating by varying the detector aperture angle $\theta$ for a full sweep of sample  orientations ($-90^\circ$ < $\alpha$ < $90^\circ$). The peak in the scattering distribution can be seen to shift as $\alpha$ is varied, matching the behaviour seen in Figure~\ref{fig:gratings_fig} (a) and (b). A plot of the results from a ray tracing simulation over the same range of $\theta$ and $\alpha$ values can be seen in (f), well matched to the experimental results.}
\label{fig:exp_vs_model_fig}
\end{figure}

Given that the wavelength of our helium beam ($\sim $\SI{0.05}{\nano\metre}) is similar in magnitude to the spacing between atoms, most surfaces, that are not single crystals, will appear randomly rough in comparison. The adsorbate covered HOPG in the previous section showed how the helium is spread out more diffusely as it scatters from roughened surfaces. We might naively expect the scattering distribution from a typical rough surface to be a cosine distribution. Indeed, such an assumption leads to accurate flow calculations in vacuum systems~\cite{steckelmacher_knudsen_1986,steckelmacher_review_1966,steckelmacher_flow_1974,m._knudsen_kinetic_1950}. However, while the nanoscopic scattering distribution might follow a cosine, macroscopic topography can lead to deviations from a cosine distribution.

A sample with a known repeating topography below the lateral resolution of the instrument was chosen to study diffuse scattering. The AFM standard `TGZ4' from TipsNano satisfies this requirement, comprised of a $\sim$ 3\,mm\,$\times$\,3\,mm  square of trenches (grating period of \SI{3}{\micro\metre} and step height of \SI{1.5}{\micro\metre}) within a SiO$_2$ frame. Figure~\ref{fig:gratings_fig}(a) and (b) show schematic representations of the grating trenches orientated parallel ($\alpha$ = $\pm 90^\circ$) and perpendicular ($\alpha$ = $0^\circ$) to the scattering plane, respectively. Figures~\ref{fig:gratings_fig}(c) and (d) show neutral helium micrographs of TGZ4 collected with $\theta$ = $0^\circ$ (i.e.: normal to the sample surface) for both the parallel and perpendicular orientations (respectively) Finally, Figure~\ref{fig:gratings_fig}(e) shows scattering distributions as collected from the outer frame of the sample, as well as the actual grating in the two orientations.

Clear contrast is seen in the pair of micrographs shown in Figures~\ref{fig:gratings_fig}(c) and \ref{fig:gratings_fig}(d) between the outer frame and the central grating. In addition, there is a contrast inversion that occurs with rotation of the sample. While the helium signal remains constant from the outer frame, the intensity from the grating was observed to vary strongly as a function of $\alpha$; a direct demonstration of image contrast being a consequence of both the local sample topography and the instrument geometry~\cite{fahy_image_2018, eder_sub-resolution_2023}. 
The observed contrast can be understood from the scattering distribution. Previous work by Lambrick et al.~\cite{lambrick_multiple_2020} has demonstrated that scattering from a deep trench in such an orientation will result in an increased signal as compared to a flat substrate. Any helium scattered out of plane has the potential to be pushed back on plane via multiple scattering (`beaming'), resulting in the grating appearing brighter than the (comparatively featureless) outer frame. When the grating is in the perpendicular orientation a portion of the helium that impinges on the bottom of each trench is now occluded by the side walls. The increase in shadowing observed in the SHeM micrographs (via the contrast inversion) manifests as a peak shift towards the backscatter direction ($\theta$ < $0^\circ$). We attribute the small feature appearing at $60^\circ$ to a small variation in the conductance of the detector during the rotation.


Figure~\ref{fig:exp_vs_model_fig}(a) shows angular distributions measured at a range of different sample orientations between the parallel and perpendicular orientations and combined to form a 2D matrix represented as an image. Figure~\ref{fig:exp_vs_model_fig}(b) is a simulation of the same procedure, repeated at each sample orientation. Bespoke computational models have previously seen success in predicting observed contrast in images for a fixed sample and detector~\cite{eder_sub-resolution_2023}. Here, we have used the general ray-tracing framework of Lambrick et al.~\cite{lambrick_ray_2018}, where we construct a model of the sample consisting of a series of flat surfaces called facets. These facets each scatter with a cosine distribution centred on the surface normal, with the atoms able to scatter between the facets multiple times before escaping. The original ray tracing program presented by Lambrick et al. has subsequently been modified~\cite{lambrick_shem-ray-tracing-simulation_2023} to facilitate the simulation of the scattering distribution; instead of only monitoring when rays pass through a small detector aperture, the direction of all escaping rays are recorded. The experiment in Figure~\ref{fig:exp_vs_model_fig}(a) and the facet scattering model in Figure~\ref{fig:exp_vs_model_fig}(b) are presented by normalising to the maximum and minimum of each dataset between 0 and 1. The qualitative agreement between the figures suggests that the model is sufficient to explain the observed results. The experimental dataset is seen to reduce more quickly with polar angle, $\theta$, when compared to the model. We attribute a component of the quantitative difference to multiple scattering between the macroscopic sample and pinhole plate which is not considered in the simulation.

While we are unable to probe the individual scattering events, comparisons between experimentation and simulation provide an insight into the origin of the scattering distribution. In particular, we see that the macroscopic topography of a surface can have a significant effect on the observed distribution.

\subsection{Diffuse scattering from a technological surface}

\begin{figure}[t!] 
\centering
{\label{fig:angular_scan_exp}%
  \includegraphics[width=\columnwidth]{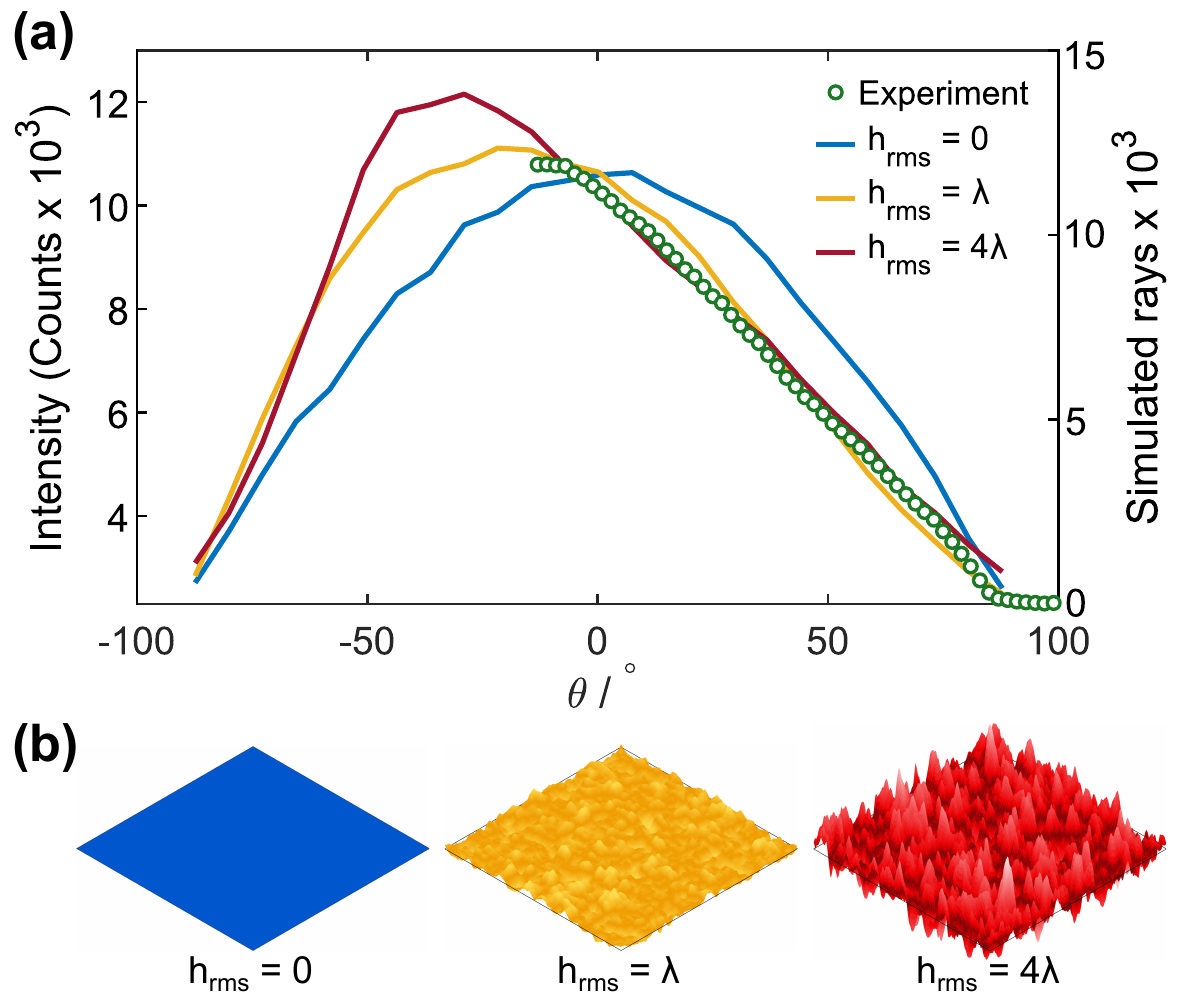}%
  }
\caption{
(a) Helium intensity as a function of detector angle $\theta$ as measured from a 3D printed resin surface plotted as green circles. Data shown is the average of 12 separate angular scans run at the same location on the sample (note that error bars are smaller than marker symbol size). The intensity approaches the maximal value beyond the surface normal ($\theta = 0$), and is likely to be outside the achievable detector angle for the stage. The solid lines correspond to ray tracing simulations from a series of random surfaces shown in (a). We observe that as the roughness is increased, the maximum in the intensity is found to shift further in the back-scattering direction.
(b) Generated randomly rough surfaces for use in ray tracing simulations, with the corresponding helium scattering distributions shown above.}
\label{fig:3d_angular_scan}
\end{figure}

Moving beyond known surface topographies, we now examine scattering from a random rough surface. A resin 3D printed surface was selected as a representative example of a technological sample, since it is not expected to posses surface order. We can still consider the surface as a series of planar facets, where the individual facets of such surfaces scatter with a cosine distribution~\cite{lambrick_observation_2022,  m._knudsen_kinetic_1950}. However, as seen with the previous sample, the aggregate scattering distribution when averaging over many facets can deviate from a cosine. These deviations arise due to two main reasons: a variation of the probability density functions of the local surface normals, and multiple scattering before escaping the sample.

Figure~\ref{fig:3d_angular_scan}(a) shows the scattering distribution obtained by an angular scan of the 3D printed test sample along with ray tracing simulations which shall be described shortly. While the distribution is broad, it does not follow $\cos\theta$. Instead, the maximum intensity is found in the backscattering direction, an effect that has been observed in the scattering of photons~\cite{smith_geometrical_1967,ginneken_diffuse_1998}, measurements of the rainbow angles of heavy atoms~\cite{moix_quantum_2011} and even directly in HAS~\cite{okeefe_atomic_1971}.

Figure~\ref{fig:3d_angular_scan}(a) also illustrates the results of ray tracing when applied to randomly rough surfaces. A random surface is generated by the procedure described in appendix~\ref{sec:random_surf} and can be characterised by the ratio of the standard deviation of surface heights, $h_{\text{rms}}$, to the correlation length of the surface, $\lambda$. Examples of these surfaces are plotted in Figure~\ref{fig:3d_angular_scan}(b). The blue surface / curve corresponds to a flat sample and exhibits the expected cosine distribution. The yellow surface / curve corresponds to a surface with $h_{\text{rms}} = \lambda$ and the red surface / curve corresponds to $h_{\text{rms}} = 4 \lambda$. We observe as the roughness is increased the maximum signal is found to shift further towards the back-scattering direction.

The preferential backscattering described above is scale invariant and only depends on the relative topography of the surface, not on the absolute length-scale. Therefore, provided the surface features are not so large they can be spatially resolved, the aspect ratio of the surface becomes a key metric to explain the observed trends in SHeM contrast data. Such an observation is consistent with the image contrast in SHeM micrographs of thin films (including 
gold island growth~\cite{eder_sub-resolution_2023} and comparisons between different metallic species~\cite{barr_unlocking_2016}). Indeed, while only data from a 3D printed resin has been presented here, we expect that the shifting of the distribution towards the backscattering direction is likely to be universal across most technological samples and therefore contributes to the contrast in most SHeM micrographs to date.

\section{Discussion}

\begin{figure}[t!]
\includegraphics[width=\columnwidth]{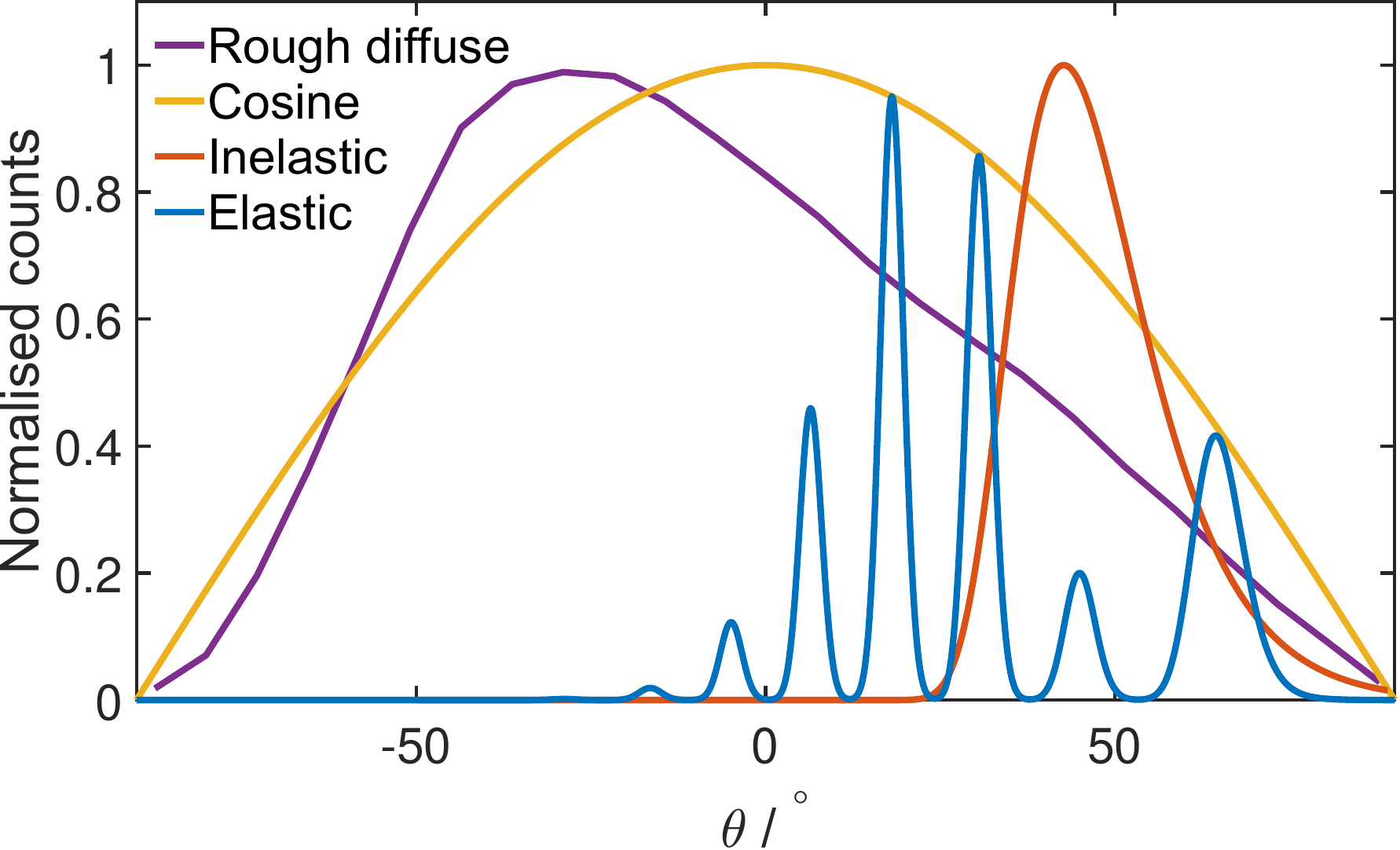}
\caption{Comparison of model scattering distributions collected exhibiting the features described in the paper. Elastic scattering from an ordered surface (such as that seen on LiF) results in series of sharp diffraction peaks. Inelastic scattering (such as that seen on HOPG) results in the diffraction pattern broadening around the specular direction. While perfectly diffuse scattering gives a cosine, diffuse scattering from a rough topography results in enhanced intensity in the backscatter direction.
}
\label{fig:comparison_fig}
\end{figure}

Given that helium atom scattering involves no surface penetration, charge interaction or chemical reaction, it might be expected that there is little contrast available from the helium surface interaction. However, the four cases presented in this paper demonstrate that in actual fact a large amount of information is encoded into the helium scattering distribution and provides the origin for the observed contrast. Figure~\ref{fig:comparison_fig} presents model scattering distributions corresponding to the samples presented earlier. Beginning with LiF, elastic diffraction from the surface results in a series of sharp peaks. Diffraction from HOPG adds in the additional complexity of inelastic scattering which leads to a broadening of the scattering distribution around the specular peak. Finally, we turn to diffuse scattering from a rough sample, where we understand that macroscopic topography can cause the pattern to deviate from cosine scattering and enhance the scattering into the backscatter direction.

We highlight the strong change in the scattering when HOPG is covered with an adsorbate layer that was shown in figure~\ref{fig:HOPG}(a). The low energy of the helium beam makes it uniquely sensitive to the electron density at the very surface of the sample. Despite only a minor change in the surface due to adsorbate contamination, the surface order is destroyed leading to a clear change in the helium scattering distribution. The scattering pattern with an adsorbate layer is a broad diffuse distribution, however the scattering cannot be totally random since cosine scattering is not observed and the maximum intensity is observed close to specular. Such a case study is indicative of the strength of scanning helium microscopy in measuring order in thin layers of materials which can now be applied to systems such as 2D materials.

\section{Conclusions}

A design for an angularly resolved detection stage for scanning helium microscopy has been presented, facilitating rotation of the sample, variation of the outgoing detector angle, and sample heating. In addition to removing previous restrictions on the scattering geometry specifically related to imaging, the design now allows for the scattering distribution from a point on the sample surface to be collected. Four examples of scattering distributions are presented that demonstate: elastic diffraction, inelastic scattering, the role of macroscopic topoghraphy in diffuse scattering and finally diffuse backscattering from a real rough surface. The addition of the angular stage highlights SHeM's extreme sensitivity to the atomic structure of surfaces, and opens up new possibilities for imaging and spectroscopy.


\section*{Acknowledgements}
The work was performed in part at the Materials and NSW node of the Australian National Fabrication Facility, a company established under the National Collaborative Research Infrastructure Strategy to provide nano and microfabrication facilities for Australia’s researchers. We would like to thank Prof Jardine, Prof Ellis and Dr Eder for useful discussions and Prof Alexandrowicz for providing the LiF sample.

%


\appendix

\section{General expression for mask length}

\label{sec:mask_length_maths}

\begin{figure}[t]
\includegraphics[width=\linewidth]{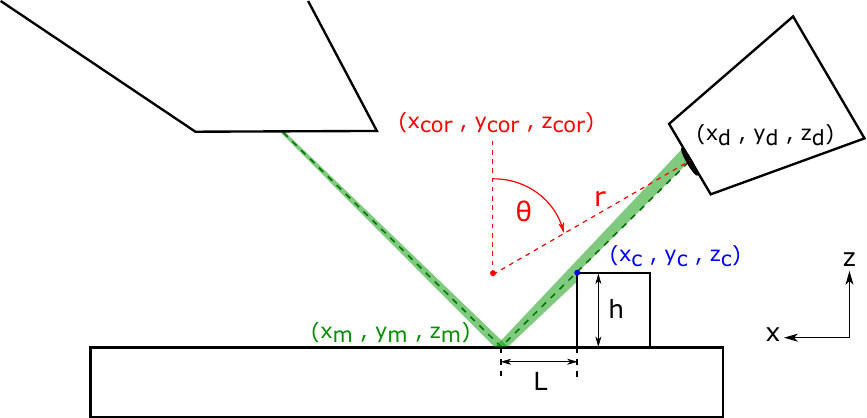}
\caption{Diagram showing the geometry of mask production with a misalignment between centre of rotation of the detector aperture arm and the scattering point on the sample.}
\label{fig:scattering_schematic}
\end{figure}

Masking occurs when helium that scatters from a sample surface is blocked from entering the detector aperture by a sample feature. Here we derive a general expression for the length of a mask created by a step feature as shown in Fig.~\ref{fig:scattering_schematic}. The point on the lower flat surface where masking begins \((x_{m}, y_{m}, z_{m})\) depends on the positions of both the detector aperture \((x_{d}, y_{d}, z_{d})\) and the corner of the step feature \((x_{c}, y_{c}, z_{c})\).

The equation of a line, $\vec{r}$, in 3D where \(\vec{a}\) and \(\vec{b}\) are points along that line is given by,
\begin{equation}
\vec{r} = \vec{a} + t \cdot ( \vec{b} - \vec{a} ).
\end{equation}
Let us consider a line projecting from the detector, through the corner of the step and down to the scattering point using the coordinates given in Fig.~\ref{fig:scattering_schematic}. By focusing on just the z-component we find the slope of the line is given by,
%
%
\begin{equation}
t = \frac{z_m-z_d}{z_c-z_d}.
\end{equation}
Substituting the slope into the x-component of the line equation gives,
\begin{equation}
x_m = x_d + \frac{z_m-z_d}{z_c-z_d} \cdot ( x_m - L - x_d ),
\end{equation}
where \(x_m - L\) is substituted for the position of the corner in the x-axis, \(x_c\). The equation can then be rearranged to find the mask length, 
%
%
%

\begin{equation}
L = h \cdot \frac{x_m - x_d}{z_m-z_d},
\end{equation}
 where $h = (z_m - z_c)$ is the height of the step \(h\). The position of the detector changes in both the x and z-axis and can be written as,
\begin{gather}
x_d = x_{cor} + r \; \sin(\theta),\\
z_d = z_{cor} + r \; \cos(\theta),
\end{gather}
where \(r\) is the radius of the detector aperture to the centre of rotation and \(\theta\) is the the angle of rotation of the detector aperture. Therefore we find, 
\begin{align}
L &= h \cdot \frac{x_m - x_{cor} + r \; \sin(\theta)}{z_m - z_{cor} + r \; \cos(\theta)},\\
&= h \cdot \frac{\alpha - \sin(\theta)}{\beta - \cos(\theta)}.
\label{eq:mask_L}
\end{align}
Where we have used,
\begin{gather}
\alpha = \frac{x_m - x_{cor}}{r} = \frac{\Delta x}{r},\\
\beta = \frac{z_m - z_{cor}}{r} = \frac{\Delta z}{r}.
\end{gather}

\section{Demonstration of detector alignment}

\label{sec:alignment}

To align the angular detector stage, the length of a mask produced by a step feature of known height is used to first determine and then correct misalignment between the centre of rotation of the detector aperture arm  and the helium beam scattering point on the sample. In Appendix~\ref{sec:mask_length_maths}, it is shown that when the stage is misaligned, the mask length is given by equation~\ref{eq:mask_L}, but when alignment is achieved, the function reduces to $L = h \; \tan(\theta)$.

\begin{figure}[t]
\includegraphics[width=\linewidth]{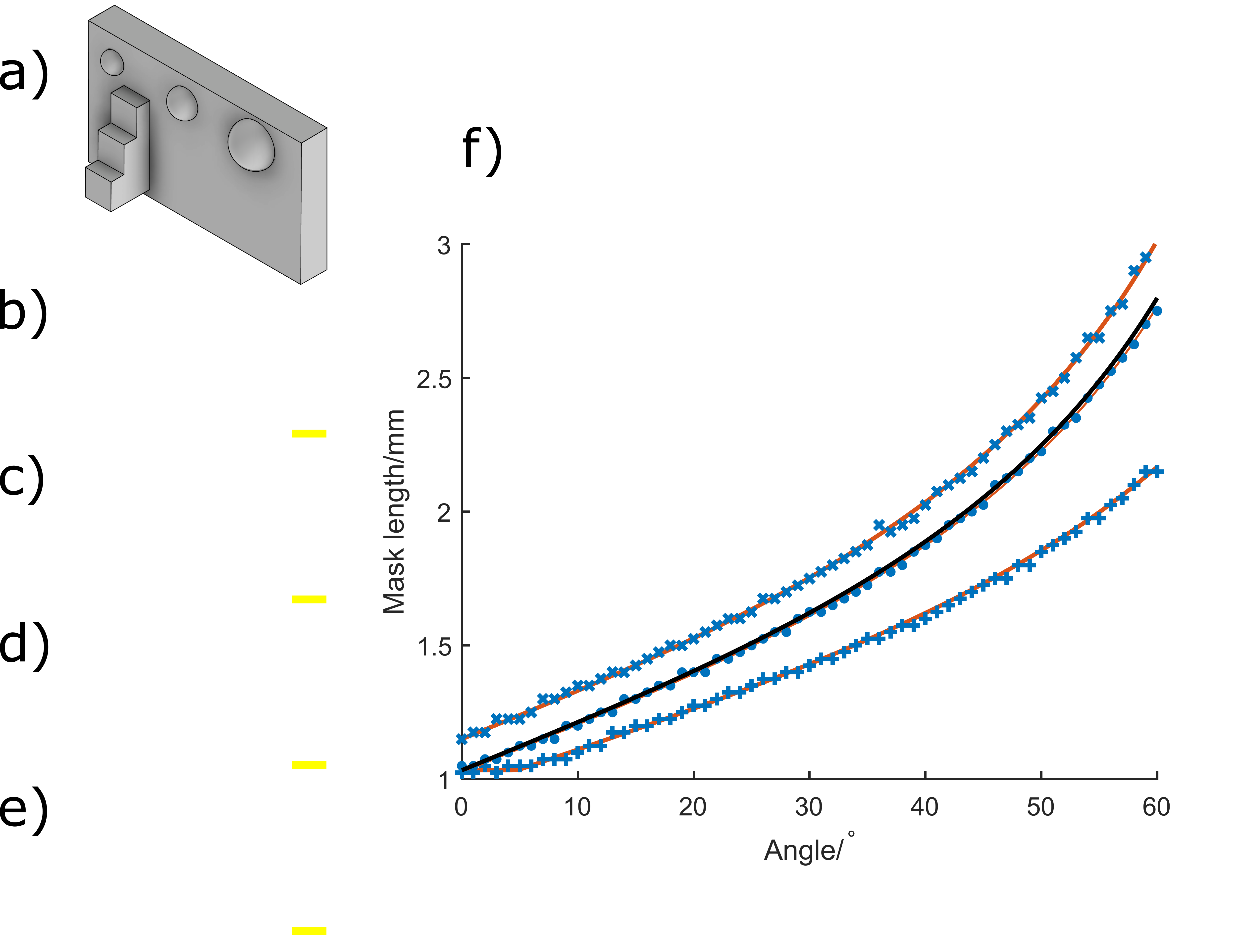}
\caption{(a) Model of the step and hemisphere test sample. SHeM micrographs of the sample taken with detector angle (b) 0°, (c) 15°, (d) 30° and (e) 45°. Scale bars are 2\,mm. (f) Simultaneously fitting the mask lengths for angular-line scans on 1\,mm step with sample z=-2.5\,mm and arm x=0\,mm plotted as +, sample z=-3.8\,mm and arm x=0\,mm plotted as x and sample z=-3.8\,mm and arm x=0.8\,mm plotted as circles. The black line indicates the ideal $\tan\theta$ curve for a fitted step height of 1.01\,mm. The orange curves indicate the fitted model for each of the positions stated above and a good agreement is seen with the experiment.}
\label{fig:Micrograph_panel}
\end{figure}

The chosen sample, shown in Fig~\ref{fig:Micrograph_panel}(a), was 3D printed out of clear V4 resin using the same process previously outlined for the stage components. The test sample has three 2\,x\,2\,mm square steps of heights 1\,mm, 2\,mm and 3\,mm for measuring the mask length and three hemispheres of diameter 1.5\,mm, 2\,mm and 3\,mm to illustrate how contrast varies across a curved surface. Micrographs of the test sample can be seen in Fig~\ref{fig:Micrograph_panel}, taken with detector aperture angles of (b) 0°, (c) 15°, (d) 30° and (e) 45°. When the detector is positioned at the surface normal as shown in (b), no mask is produced as the detector aperture has a direct line of sight to the base of the steps. As the detector angle shifts and line of sight is increasingly blocked by the side of the step, the mask length grows.

To measure how the mask length varies with detector angle, a series of line scans that ran across the lower flat region, up the side wall and on top of a chosen step were acquired. The line scans were repeated for a range of the detector angles between the surface normal 0° to a maximum of 70°.

Figure~\ref{fig:Micrograph_panel}(f) shows how the mask length changes with detector angle on the nominally 1\,mm step at various z-axis positions. All data sets were simultaneously fitted using Eqn~\ref{eq:mask_L} to determine the misalignment between the scattering point and the detector aperture centre of rotation. A good agreement is observed between the model and the experimental data for each position of the detector. By carefully choosing both the sample $z$ position and the $x$ position of the detector aperture arm, we are able to ensure that the mask length closely follows $\tan\theta$ (plotted in black) and therefore the detector is aligned with the sample scattering point.

As the detector is aligned, we have confidence that the acceptance aperture may be rotated without varying the solid angle during measurement and that the angle of the detector matches the true outgoing angle of the helium from the surface. Therefore, a direct measurement of the helium scattering distribution of the helium from a sample is possible.

\section{Random surface generation}

\label{sec:random_surf}

A discrete random surface, $f_n$, is generated with a known coherence length, $\lambda$, by taking a convolution of points sampled from a Gaussian with an exponential~\cite{bergstrom_absorption_2008,garcia_monte_1984},
\begin{equation}
    f_{n,m} = \sum_{i,j} h Z_{i,j} e^{-|n-i|\Delta x / \lambda}\, e^{-|m-j|\Delta x / \lambda},
    \label{eq:surface_sum}
\end{equation}
where $\Delta x$ is the point spacing, $Z_{i,j}$ is a random variable sampled from a Gaussian $\mathcal{N}(\mu =0, \sigma^2)$. The above description is numerically implemented using the convolution theorem and multiplying the Fourier transforms of the Gaussian variable and exponentials.

The root mean square and coherence length of the height of the surface can both be determined by first calculating the expectation value $\left\langle f_{n,m} \, f_{n+a,m+b} \right\rangle$. Since the Gaussians $Z_{i,j}$ are independent, we have that $\left\langle Z_{i,j} \, Z_{k,l} \right\rangle = \sigma^2 \delta_{i,k} \delta_{j,l}$, the cross terms disappear and we obtain,
\begin{multline}
   \left\langle f_{n,m} \, f_{n+a,m+b} \right\rangle = h^2 \sigma^2 \sum_{i,j} \biggl[ e^{-|n-i|\Delta x / \lambda}\, e^{-|n+a-i|\Delta x / \lambda} \\ e^{-|m-j|\Delta x / \lambda}\, e^{-|m+b-j|\Delta x / \lambda} \biggr].
\end{multline}
The resultant sum is separable and we begin by solving the sum in each dimension individually. For $a>0$, if we assume an infinitely large surface, it can be shown that,
\begin{align}
    S  &= \smashoperator{\sum_{i=-\infty}^{\infty}}  e^{-|n-i|\Delta x / \lambda}\, e^{-|n+a-i|\Delta x / \lambda}\\
    & =  \smashoperator{\sum_{i=-\infty}^{n-1}} e^{(-2n-a+2i)\Delta x / \lambda} + \smashoperator{\sum_{i=n}^{n+a}} e^{-a\Delta x / \lambda}
    + \smashoperator{\sum_{i=n+a+1}^{\infty}} e^{(2n+a-2i)\Delta x / \lambda}. 
\end{align}
Considering the first and last sums as geometric series and repeating a similar procedure for $a<0$, we obtain the general expression,
\begin{equation}
    S =  \frac{2e^{-(|a|+2)\Delta x / \lambda}}{1-e^{-2\Delta x / \lambda}} + (|a|+1)e^{-|a|\Delta x / \lambda}
\end{equation}
Rearranging the above equation obtains,
\begin{equation}
    S = e^{-|a|\Delta x / \lambda}\left[|a| + \coth \left(\frac{\Delta x}{\lambda}\right)\right].
\end{equation}
We can now show that the expectation value is given by,
\begin{multline}
   \left\langle f_{n,m} \, f_{n+a,m+b} \right\rangle = h^2 \sigma^2 e^{-|a|\Delta x / \lambda} e^{-|b|\Delta x / \lambda}\\ \left[|a| + \coth \left(\frac{\Delta x}{\lambda}\right)\right] \left[|b| + \coth \left(\frac{\Delta x}{\lambda}\right)\right].
\end{multline}
We therefore find that the root mean square of the surface heights, $h_{\text{rms}} = \left\langle f_{n,m} \, f_{n,m} \right\rangle^{1/2}$ in the limit of an infinite number of points is,
\begin{equation}
    h_{\text{rms}}=h \sigma \coth \left(\frac{\Delta x}{\lambda}\right)
\end{equation}
It can also be shown that the normalised autocorrelation function, $\rho_{a,b} = \left\langle f_{n,m} \, f_{n+a,m+b} \right\rangle / \left\langle f_{n,m} \, f_{n,m} \right\rangle $, for the random surface is given by,
\begin{equation}
\rho_{a,b} = e^{-(|a|+|b|)\Delta x/\lambda} \left(1+\frac{|a|}{\coth \left(\frac{\Delta x}{\lambda}\right)}\right) \left(1+\frac{|b|}{\coth \left(\frac{\Delta x}{\lambda}\right)}\right).    
\end{equation}
Therefore, the coherence length of the surface can be approximated to be $\lambda$, since the exponential term at the start dominates the behaviour.

\end{document}